\documentclass[AMA,Times1COL]{WileyNJDv5} 
\articletype{Research Paper}%

\usepackage{graphicx}

\usepackage{multirow}
\usepackage{booktabs}
\usepackage{bigstrut}
\usepackage{blkarray}

\usepackage{physics}
\usepackage{microtype}
\usepackage{xcolor}
\usepackage{orcidlink}
\hypersetup{
    colorlinks=true,
    linkcolor=blue,
    citecolor=blue,
    urlcolor=blue
}
\startpage{1}

\begin{document}

\title{Detecting Mutations in double-stranded DNA using Graphene nanopores}

\author[1]{Sourav Kundu}
\author[1]{Siddhartha Lal}

\authormark{Kundu \textsc{et al.}}
\titlemark{Detecting Mutations in double-stranded DNA using Graphene nanopores}

\address[1]{%
  \orgdiv{Department of Physical Sciences}, 
  \orgname{Indian Institute of Science Education and Research Kolkata}, 
  \orgaddress{\state{Mohanpur, West Bengal - 741246}, \country{India}}%
}


\corres{Sourav Kundu. Department of Physical Sciences, Indian Institute of Science Education and Research Kolkata, Mohanpur, West Bengal - 741246, India. \email{sourav.kunduphy@gmail.com}}

\abstract[Abstract]{Graphene nanopores have primarily been used for single-molecule DNA sequencing, overlooking issues arising from mutations in DNA that can impact healthcare outcomes. To address this concern, we conduct a study using a tight-binding formulation to analyze a two-terminal graphene nanopore device for double-stranded DNA sequencing. By employing the non-equilibrium Green's function (NEGF) technique, we investigate the changes in electronic transport properties as double-stranded DNA traverses a nanopore cavity embedded within a zigzag graphene nanoribbon. We first characterize the device under static conditions in order to optimise the ability to distinguish mismatches from canonical basepairs. We then apply an optimal voltage bias for sequencing and recording the corresponding current signals for each base-pair of the translocating DNA. The study successfully identifies standard AT \& GC pairs, as well as various base-pair mismatches, including methylated pairs. Furthermore, we explore different edge atom functionalization of the nanopore in order to improve device efficiency. Our results offer a first glimpse on transport through graphene nanopores as effective detectors of mutated basepairs.}


\maketitle

\section{Introduction}
\begin{figure*}[htb]
\centering
\includegraphics[width=0.5\textwidth, bb=0 0 2000 1400]{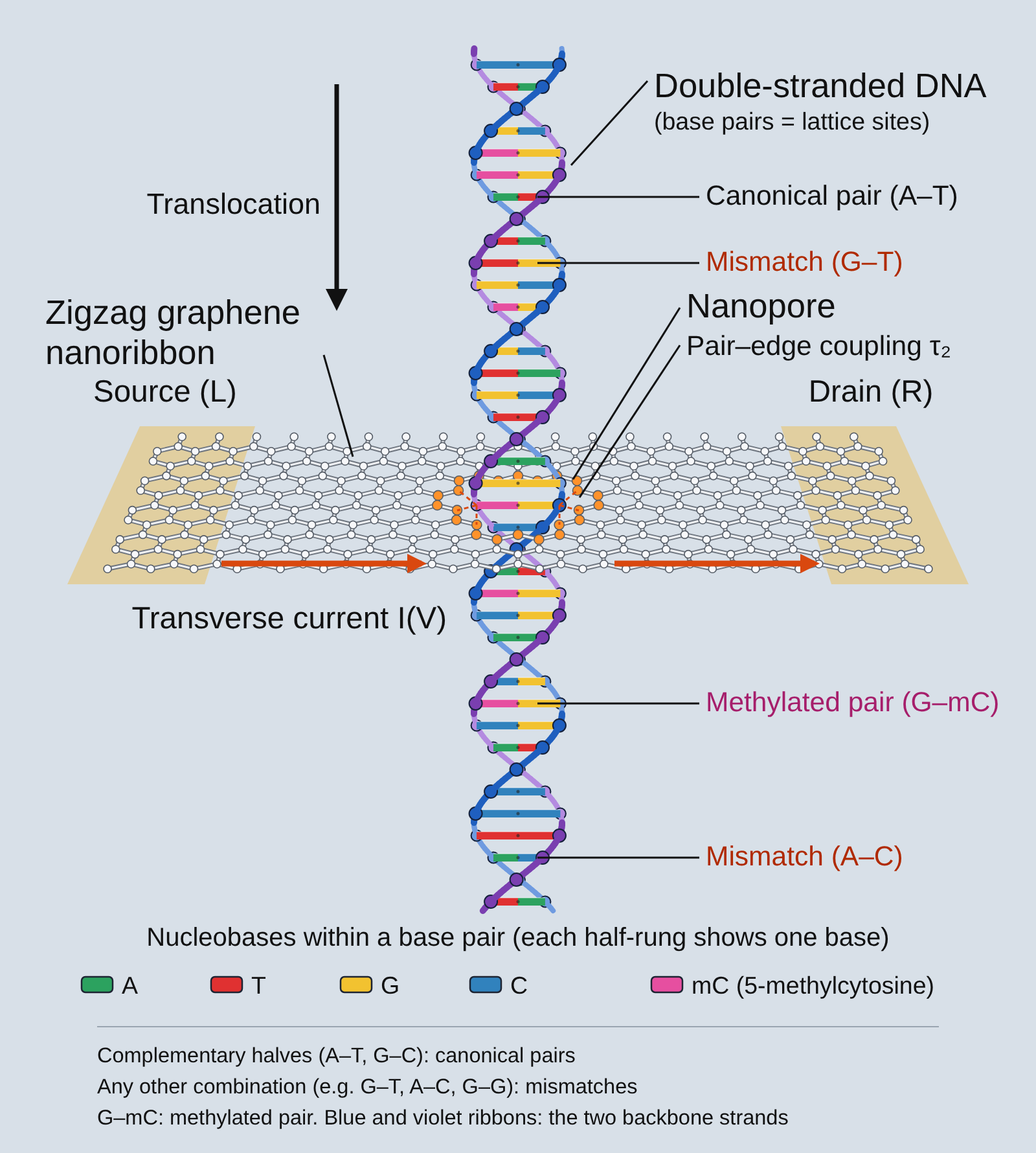}	
\caption{(Color online). Schematic view of a graphene nanopore device with a cavity through which a double stranded DNA is passed. Current-voltage characteristics are studied by applying a voltage difference between the source and drain leads, and measuring the current that flows due to the bias.}
\label{fig1}
\end{figure*}

DNA is the fundamental building block of life, and is comprised of three major parts: nitrogen bases, pentose sugar, and phosphate groups. The pentose sugar and phosphate groups form the backbone structure of DNA, while the nitrogen bases (connected to one another via hydrogen bonds) form the ladder-like structure known as the DNA double helix~\cite{watson}. There are four nitrogen bases: Adenine (A), Thymine (T), Guanine (G), and Cytosine (C). Adenine pairs with Thymine, and Guanine pairs with Cytosine, known as complementary base-pairing. Any other combination of these bases is considered a mismatch and can lead to various epigenetic disorders in living organisms. Fluorescence-based techniques (e.g., the Sanger~\cite{sanger} method) that are both costly and time-consuming have traditionally been employed for DNA sequencing. New DNA sequencing techniques, such as single-molecule electrical analysis using nanopores, single-molecule junctions (SMJ), and single-molecule field-effect transistors (FET) offer the real-time monitoring of molecular structure changes through current fluctuations~\cite{Wu_talanta,Motone_nature,Brinker_science,Wang_pnas}. Among these, graphene nanopore devices have emerged as promising tools for DNA sequencing at the single-molecule level~\cite{sourav1,ralph1,pathak1,pathak2,pathak3}. These devices utilize properties that are unique to graphene, such as its high electrical conductivity~\cite{rocha_book} and single-atom thickness~\cite{novo1,novo2}, to detect changes in DNA sequences. By passing DNA strands through nanopores in graphene (see Fig.~\ref{fig1} for a schematic representation), researchers can identify the bases by measuring changes in electrical current as the DNA passes through the nanopore. Graphene-based nanopores offer various methods for sequential detection, including transverse tunneling~\cite{postma,prasong}, nanoribbon conductance~\cite{saha,nelson}, and mechanical deformation of DNA~\cite{sathe}. A few experimental realizations of these methods have also been reported~\cite{traversi,menard}. Other two-dimensional (2D) materials such as silicene~\cite{ralph2,qli,gurel} and graphdiyne~\cite{pathak4,rezapour} have also been explored for DNA sequencing. For more detailed information on nanopore-based sequencing techniques, readers are referred to review articles~\cite{moghadam,fyta,ralph,rashid,ventra1,pathak3}. 

Despite these advancements, however, nanopore sequencing techniques are primarily tested on single-stranded DNA~\cite{ralph1,pathak2,pathak3}, overlooking the possibility of base-pair mismatches. In this study, we investigate the detection of base-pair mismatches in double-stranded DNA using a graphene nanopore device. Instead of focusing on single-molecule detection, we aim to sequence at the base-pair level. Through theoretical analysis based on a tight-binding formulation, we explore the characteristics of canonical base pairs (AT and GC) and five different types of mismatches. By employing non-equilibrium Green's function (NEGF) techniques together with the Landauer-Buttiker formalism~\cite{datta1,datta2}, we then calculate the conductance and current outputs for seven base pairs (including five mismatches). Additionally, we analyze the local electronic density of states (LDOS) for each base by trapping them at the nanopore, and investigate thereby the functionalization of nanopore edge atoms in both the weak and strong coupling limits. Our numerical results demonstrate that not only can canonical base pairs (AT and GC) be distinguished based on their transport properties, but that different types of mismatches can also be identified. Importantly, varying the coupling strength between the nanopore edge and the DNA from weak to strong provides better resolution in distinguishing mismatches. This indicates that functionalizing the nanopore edge atoms with different atoms or molecules can further enhance the detection capabilities.
\vspace*{1.0cm}
\par
The remainder of this paper is structured as follows. We introduce the model Hamiltonian and outline the theoretical framework in Section II, and present the analysis of the numerical results in Section III. Finally, we conclude in Section IV. 

\section{Theoretical Formulation}
To determine the base-pair sequence based on its electrical properties, we employ a model of a graphene nanoribbon with zigzag edges with a pore at its center (Fig.~\ref{fig1}). The pore is designed to maintain the bipartite two-sublattice nature of graphene. Further, electrodes made of semi-infinite zigzag graphene nanoribbons (ZGNRs) are attached to both sides of the device. The active system, which includes the nanopore but excludes the electrodes, is henceforth referred to as the \textit{device} and can be described by the following tight-binding Hamiltonian (refer to Fig.~\ref{fig1}):
\begin{eqnarray}
& H_{zgnr}&= \sum\limits_{i=1}^N\left(\epsilon_{i}
c^\dagger_{i}c_{i}+t c^\dagger_{i}c_{i+1}+\mbox{H.c.} \right)~,
\label{eqzgnr}
\end{eqnarray}
where $\epsilon_{i}$ is the on-site energy of each carbon atom in ZGNR, and $t$ is the nearest neighbour hopping amplitude. Further, $c_i$ and $c^\dagger_{i}$ create and annihilate an electron at the $i$th site respectively. For the calculation of transport properties, we also employ 
semi-infinite ZGNRs as electrodes~\cite{saha}. The Hamiltonian of the ZGNR electrodes can also be expressed as eq.\eqref{eqzgnr}. Thus the total Hamiltonian 
of the system can be written as $ H_{tot}=H_{zgnr}+ H_{leads}+H_{tun}~$, where $H_{tun}$ = $\tau_1 \left(c^\dagger_0c_1+c^\dagger_Nc_{N+1} + \mbox{H.c.}\right)~$ represents the tunneling Hamiltonian between the active nanopore device and the electrodes, with $\tau_1$ as the coupling parameter. As both the device and the electrodes are ZGNR systems, we set $\tau_1 = t$. 
\par
The Hamiltonian of a double stranded-DNA can be expressed as~\cite{iguchi1997}
\begin{equation}
 H_{DNA}= \sum\limits_{i=1}^N\sum\limits_{j=I,II}\left(\epsilon_{ij}
d^\dagger_{ij}d_{ij} +t_{ij}d^\dagger_{ij}d_{i+1j}+\mbox{H.c.} \right) + \sum_{i=1}^N v \left(d^\dagger_{iI}d_{i II} + \mbox{H.c.} \right)~,
\label{dnaHam}
\end{equation} 
where $d_{ij}^\dagger$ and $d_{ij}$ are the electron creation and annihilation 
operators at the $i$th nucleotide of the $j$th strand ($j=1,2$), $t_{ij}$ the nearest neighbour hopping amplitude between nucleotides along the $j$th strand, $\epsilon_{ij}$ is the on-site energy of the nucleotides, and $v$ the inter-strand hopping between the DNA bases. The coupling between the nanopore device and the base-pairs is given by $H_{coupling}$ = $\tau_2 \left(c^\dagger_id_{ij}+d^\dagger_{ij}c_{i} + \mbox{H.c.}\right)~$, where $\tau_2$ is the (electron hopping) coupling parameter between a base-pair and the edge atoms of the nanopore.

In this two-terminal setup, we utilize the Green's function approach to compute the transmission probability $T(E)$ of electrons~\cite{datta1,datta2} through the nanopore device. The single-particle retarded Green's function operator that represents the entire system, including the nanopore device and two semi-infinite electrodes, at an energy $E$ can be expressed as:
\begin{equation}
G^\mathrm{r}=(E-H+i\eta)^{-1}~,
\end{equation}
where the parameter $\eta\rightarrow0^+$, and $H$ is the Hamiltonian of the entire system including the DNA bases. Using the Fisher-Lee relation~\cite{datta1,fisher}, the two terminal transmission probability is defined as 
\begin{equation}
T(E)={\mathrm{Tr}} [\Gamma_\mathrm{S} G^\mathrm{r} \Gamma_\mathrm{D} G^\mathrm{a}]~, 
\end{equation}
where $E$ is the incident electron energy, $G^\mathrm{a}$ is the advanced Green's function operator, and ``Tr" represents the trace over the reduced Hilbert space spanned by the ZGNR.

The effective Green's functions can be expressed in the reduced Hilbert space using the self-energies of the source and drain electrodes
\begin{equation}
G^\mathrm{r} = [G^\mathrm{a}]^\dagger = [E-H_{\mathrm{zgnr}}-\Sigma^\mathrm{r}_S-\Sigma^\mathrm{r}_D+i\eta]^{-1}~, 
\end{equation}
where the self-energy of the source (S) and drain (D) leads is given by
\begin{equation}
\Sigma^{\mathrm{r}(\mathrm{a})}_{S(\mathrm{D})} = H^\dagger_{\mathrm{tun}} G^{\mathrm{r}(\mathrm{a})}_{S(\mathrm{D)}} H_{\mathrm{tun}}~,
\end{equation}
where $G^{\mathrm{r(a)}}_{\mathrm{S}(\mathrm{D})}$ being the retarded (advanced) Green's function for the source 
(drain) electrode and $\Sigma^{\mathrm{r(a)}}_{\mathrm{S}(\mathrm{D})}$ is the retarded (advanced) self-energy of the source (drain) electrode.  
We define the coupling matrices 
\begin{equation}
\Gamma_{S(\mathrm{D})} = i[\Sigma^\mathrm{r}_{\mathrm{S}(\mathrm{D})} - \Sigma^\mathrm{a}_{\mathrm{S}(\mathrm{D})}]=-2~{\mathrm{Im}} (\Sigma^\mathrm{r}_{\mathrm{S}(\mathrm{D})})~, 
\end{equation}
where ``Im" represents the imaginary part of the argument. The self-energies $\Sigma^\mathrm{r}_{\mathrm{S}(\mathrm{D})}$ can be written as
\begin{equation}
\Sigma^\mathrm{r}_{\mathrm{S}(\mathrm{D})}=
\Delta_{\mathrm{S(D)}} + i \Lambda_{\mathrm{S(D)}}~, 
\end{equation}
where $\Delta_{\mathrm{S(D)}}$ is the real part of $\Sigma^\mathrm{r}_{\mathrm{S(D)}}$ and corresponds to the shift of the energy levels of DNA, while the imaginary part $\Lambda_{\mathrm{S(D)}}$ leads to the broadening of these levels.
\par
Now, the current $I$ passing through the system for an applied bias voltage $V$ can be written as 
\begin{equation}
I(V) = \frac{2e}{h} 
\int^{E_\mathrm{F} + eV/2}_{E_\mathrm{F}-eV/2} T(E)~dE~, 
\end{equation}
where the Fermi energy $E_\mathrm{F}$ is set at $E_\mathrm{F}$=0 eV. The two-terminal Landauer conductance at $T=0$ is then given by 
\begin{equation}
g = \frac{2e^2}{h}T(E=E_\mathrm{F})~. 
\end{equation}

The electronic density of states (DOS) of the system can also be obtained using the Green's function formalism~\cite{sourav3}. The local density of states (LDOS) profiles of the basepairs trapped inside the nanopore are given by 
\begin{equation}
\rho(E,i) = - \frac{1}{\pi} {\rm Im[G_{ii}(E)]}~,
\end{equation}
where $G(E)= (E-H+i\eta)^{-1}$ is the Green's function for the ZGNR system (including the base-pairs) with electron energy $E$ as $\eta\rightarrow0^+$ and $H$ is the Hamiltonian of the ZGNR-nanopore system.
The local Green's function $G_{ii}(E)$ is the diagonal matrix element $(< i|G(E)|i >)$ of the full Green’s function, with $|i >$ being the Wannier state associated with the trapped nucleotide.

\section{Results and Discussions}
For the numerical calculations, we set the hopping parameter $t$ = 1.0 eV and the on-site energy $\epsilon_i$ = 0 eV for the ZGNR nanopore device and the electrodes. The ionization potentials of the nitrogen bases are used as their site energies; these are extracted from the {\it ab-initio} calculations of Ref.\cite{senth}: $\epsilon_G$ = 8.178 eV, $\epsilon_A$ = 8.631 eV, $\epsilon_C$ = 9.722 eV, and $\epsilon_T$ = 9.464 eV. Following Ref.\cite{yu_peng}, the on-site energy fo the 5-methyl-Cytosine (mC) is taken as $\epsilon_{mC}$ = 8.78 eV. In what follows, we shift the reference point of energy to the average of the ionization potentials of the bases, which is $\epsilon_{avg}$ = 8.995 eV. With respect to this new origin of energy, the on-site energies for bases G, A, C, T and methylated cytosine (mC) become -0.82 eV, -0.37 eV, 0.72 eV, 0.47 eV, and -0.22 eV respectively. Consistent with earlier works~\cite{paez,wolf,sourav3} where the average of the ionization potential was set as the backbone site energy, the transport behavior of the system is not affected by this shift of the origin of energy. In addition to the regular AT and GC pairs, we have selected five different base pair mismatches for sequential detection. Along with GG, CC, AA and TT, this includes the methylated-GC pair (GmC): we note that the methylation of nitrogen bases is often overlooked, but can have significant implications in gene regulation. The inter-strand hopping $\nu$ in eq.\eqref{dnaHam} is taken as 0.047 eV between A and T, and as 0.055 eV between G and C~\cite{senth}. For
a mutated pair, the inter-strand hopping is taken as the average of the canonical A-T and G-C pair hydrogen-bond strengths: $\nu$ = $(0.055 + 0.047)/2$ = $0.051$ eV.~\cite{souravCJP2}

\begin{figure*}[htb]
\centering
\begin{tabular}{c}
\includegraphics[width=0.42\textwidth]{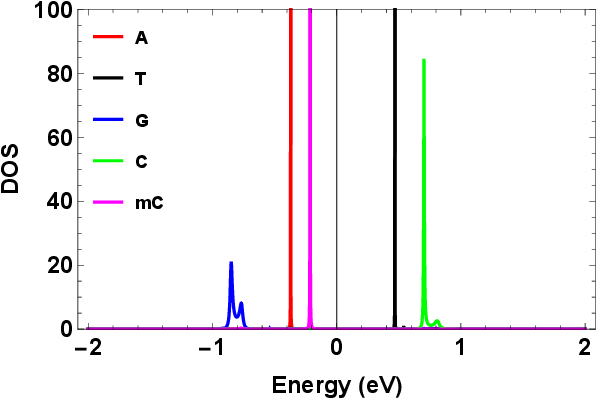}	
\end{tabular}
\caption{(Color online). Variation in the local density of states (DOS) of the four nucleotides and methylated cytosine (mC) trapped at the nanopore. Five distinct peaks of varying heights correspond to different bases near their characteristic site energies.}
\label{fig2}
\end{figure*}

In Fig.~\ref{fig2}, we display the LDOS profiles for the five different nitrogen bases A, T, G, C and methylated Cytosine (mC). Importantly, we analyze the LDOS of these bases within basepairs rather than individually. Specifically, we trap the AT, GC, and GmC pairs inside the nanopore and examine the LDOS profiles of the respective bases. The positions of the LDOS peaks are close to the characteristic site energies of the different nucleotides. The differences in LDOS values of the five nitrogen bases offer an opportunity to detect the base-pairs using AFM or STM techniques by trapping them inside the nanopore. Since the LDOS behavior is primarily influenced by the nitrogen bases rather than the backbones~\cite{he}, this approach also introduces a novel method for biomolecular detection.

\begin{figure*}[htb]
\centering
\begin{tabular}{cc}
\includegraphics[width=0.42\textwidth]{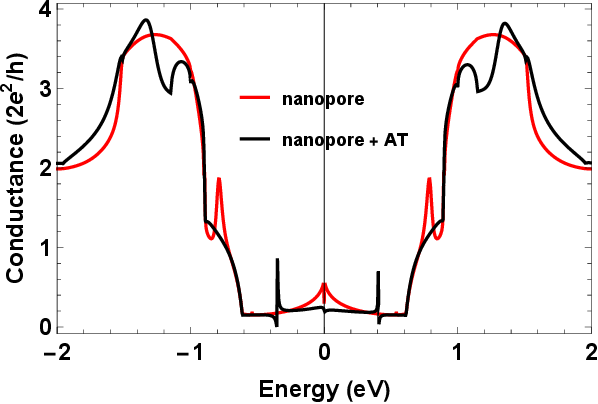}
\includegraphics[width=0.42\textwidth]{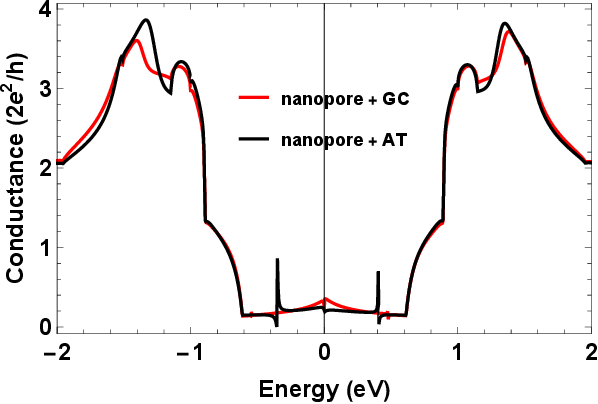}\\
\includegraphics[width=0.42\textwidth]{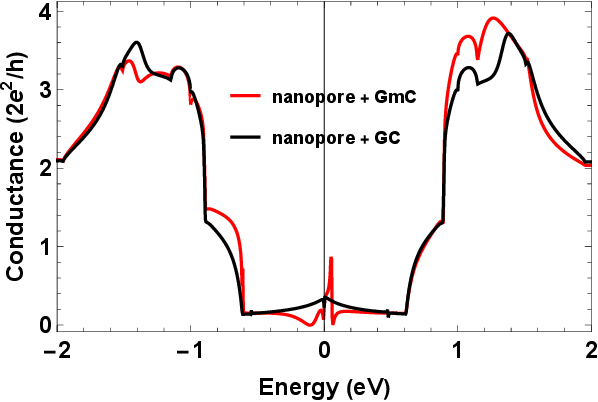}	
\includegraphics[width=0.42\textwidth]{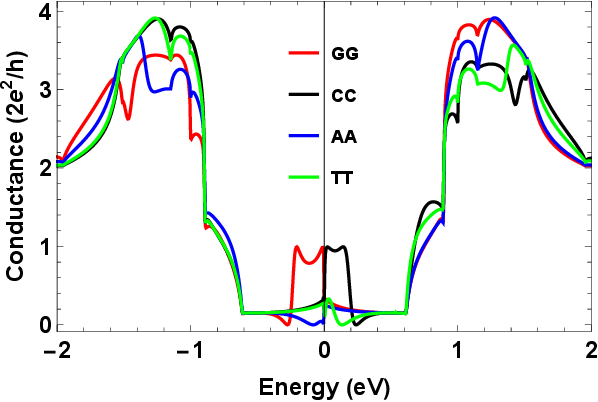}	
\end{tabular}
\caption{(Color online). Conductance as a function of energy for various scenarios of the strongly coupled nanopore-DNA system. Upper left: comparison between a bare nanopore and an AT-nanopore. Upper right: comparison between canonical pairs AT and GC. Lower left: comparison between GC and methylated-GC (GmC) pairs. Lower right: comparison of four different base-pair mismatches trapped inside the nanopore. See main text for discussion.}
\label{fig3}
\end{figure*}

\begin{figure*}[htb]
\centering
\begin{tabular}{cc}
\includegraphics[width=0.42\textwidth]{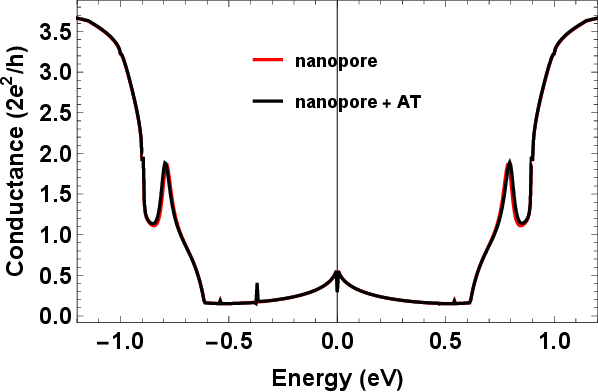}
\includegraphics[width=0.42\textwidth]{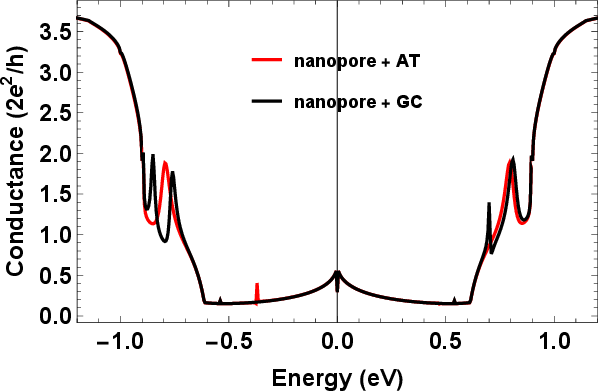}\\
\includegraphics[width=0.42\textwidth]{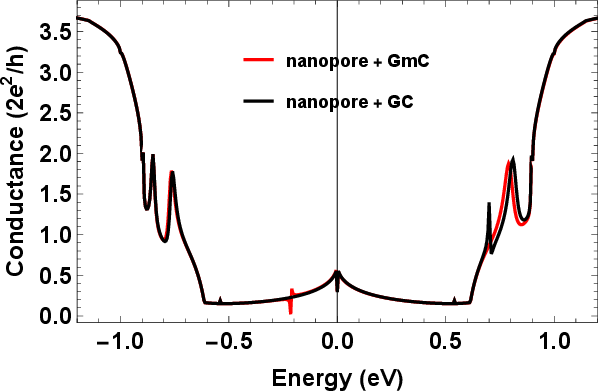}	
\includegraphics[width=0.42\textwidth]{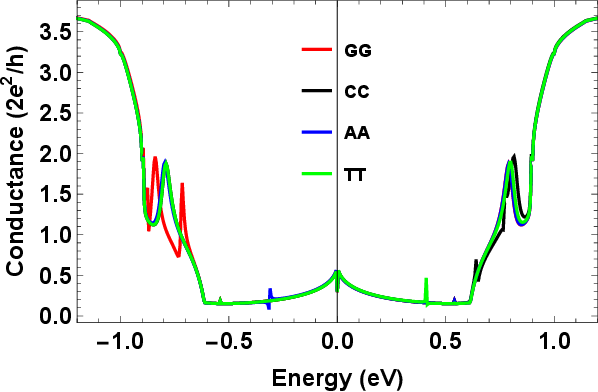}	
\end{tabular}
\caption{(Color online). Conductance as a function of energy for various scenarios of the weakly coupled nanopore-DNA system. Upper left: comparison between a bare nanopore and an AT-nanopore. Upper right: comparison between canonical pairs AT and GC. Lower left: comparison between GC and methylated-GC (GmC) pairs. Lower right: comparison of four different base-pair mismatches trapped inside the nanopore. See main text for discussion.}
\label{fig4}
\end{figure*}

In Fig.~\ref{fig3} and Fig.~\ref{fig4}, we plot the variation in conductance for different base pairs, including the five mismatches mentioned above. These two figures correspond to two types of coupling that have been considered between the boundary sites of the ZGNR-nanopore and DNA base pairs respectively, mimicking two types of functionalization: Fig.\ref{fig4} corresponds to a strong coupling with $\tau_2$ = 0.5 eV, while Fig.\ref{fig3} corresponds to a weak coupling with $\tau_2$ = 0.05 eV. We note that while the conductance versus energy curve for an empty nanopore cavity is symmetric with respect to zero energy due to the two-sublattice symmetry of the graphene nanopore system, the asymmetry observed in the curves for the cases when various basepairs are introduced within the nanopore cavity arises from the differences in their on-site and inter-strand hopping energies.
\par
First, we present the results for the strong coupling case ($\tau_2$ = 0.5 eV). The upper left plot of Fig.\ref{fig3} shows the comparison between a bare nanopore and an AT pair trapped in the nanopore. The changes in transmission spectra are clearly distinguishable, with characteristic peaks in the profile at both positive and negative energy ranges.  The upper right figure illustrates the difference between the characteristic features of the two canonical basepairs AT and GC, with distinct peaks in the transmission profile noticeable at specific energies once again. Similarly, the variations of the conductance profiles of the GC and GmC pairs are shown in lower left figure. The methylated pair is visibly separable from GC, with a distinct peak very close to zero energy; as methylation is sometimes overlooked in nanopore sequencing, such a distinctive feature in the transport can aid in distinguishing methylated basepairs. Finally, in the lower right figure, the conductance profiles for four different types of base pair mismatches (GG, AA, CC and TT) are displayed: the curves reveal a clear contrast at both low and high energies with respect to the AT and GC basepairs (upper right plot). For instance, the CC mismatch exhibits a distinct peak within energies in the range between 0 to 0.2 eV, while the rest are clearly distinguishable via features lying at energies $\sim$ 0.9 eV and $\sim$ 1.5 eV. These variations reflect the difference in their electronic structure.

In Fig.~\ref{fig4}, we present the conductance versus energy plots variations in the same sequence as in Fig.~\ref{fig3}, but for a weak coupling between ZGNR boundary and the basepair ($\tau$=0.05 eV). It is evident that with a significant decrease in coupling, the distinguishable features are diminished in all the plots, with subtle changes visible at specific energies. This reduction in features is expected as the coupling parameter connects the base pairs to the nanopore, and renders it difficult to discern different basepairs in such weak coupling conditions. The contrast between strong and weak coupling between the nanopore edge atoms and the basepairs thus reveals the importance of the functionalisation of the edge atoms in governing the effectiveness of basepair mismatch detection in such a device. This can be achieved using various agents, ranging from inorganic atoms to organic macromolecules~\cite{pathak2, pathak5}. This feature is unique to graphene and other 2D nanopores that utilize transverse conductance for electrical detection of biomolecules, unlike other nanopore sequencing techniques based on ionic blockade current~\cite{jeffery, chen2017, carral2021, chen2024}. In the remainder, therefore, we focus on the properties of a strong coupling based device.

\begin{figure*}[htb]
\centering
\begin{tabular}{cc}
\includegraphics[width=0.42\textwidth]{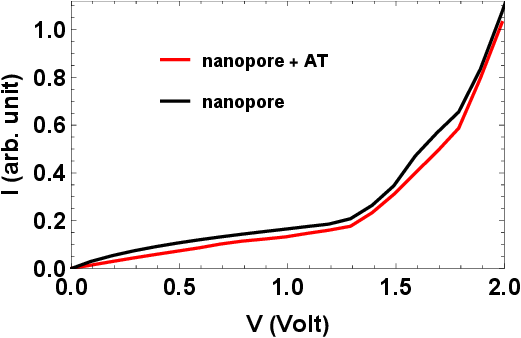}
\includegraphics[width=0.42\textwidth]{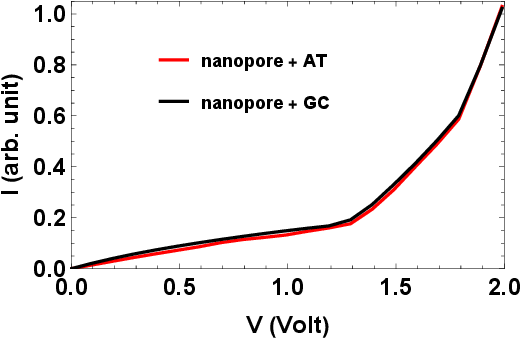}\\
\includegraphics[width=0.42\textwidth]{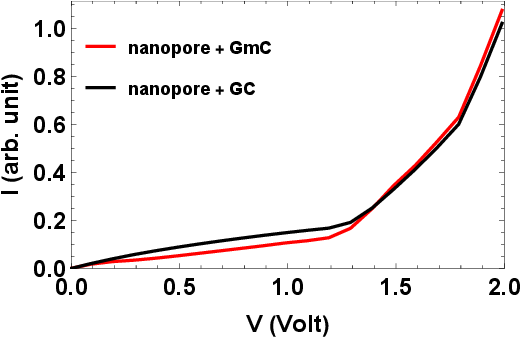}	
\includegraphics[width=0.42\textwidth]{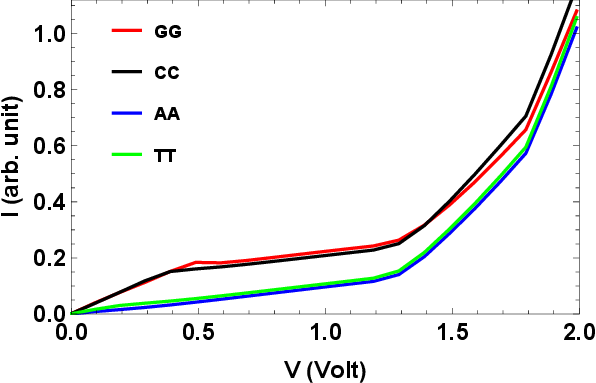}	
\end{tabular}
\caption{(Color online). Current-voltage response of the strongly coupled nanopore device for various scenarios. Upper left: comparison of current responses between a bare nanopore and an AT-nanopore. Upper right: comparison of the two canonical base pairs AT and GC. Lower left: comparison of GC and methylated-GC (GmC) pairs. Lower right: comparison of four different mismatches (GG, AA, CC, TT). See main text for discussion.}
\label{fig5}
\end{figure*}

\begin{figure*}[htb]
\centering
\begin{tabular}{c}
\includegraphics[width=0.42\textwidth]{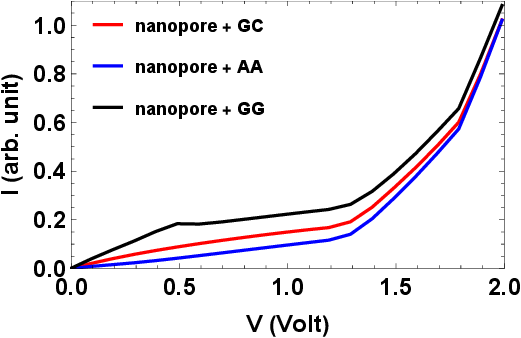}	
\end{tabular}
\caption{(Color online). Comparison of current response versus applied voltage for GC, AA and GG basepairs trapped in the nanopore. The AA and GG mismatches can clearly be distinguished from the canonical basepair GC.}
\label{fig6}
\end{figure*}

In Fig.~\ref{fig5}, we illustrate changes in the I-V characteristics for different basepair cases in the strong coupling limit, and in the same sequence as displayed in Fig.\ref{fig3}. As shown in the upper left plot, the current through the device in the presence of an AT pair within the nanopore is consistently lower (across the entire bias voltage range up to 2 Volts) than that for the empty nanopore cavity, and should therefore be possible to detect. A similar conclusion can be reached for the GC and GmC basepairs comparison shown on the lower left plot. The upper right figure reveals a much smaller separation between AT and GC basepairs, and will need a greater current sensitivity in order to distinguish between them. However, the lower right figure reveals that while the I-V characteristics of the GG and CC mutation basepairs are very close to one another, as are the AA and TT mutation basepairs, the two pairs of mutations are well separated from one another. Importantly, as shown in Fig.~\ref{fig6}, these two pairs can be distinguished from the AT and GC basepairs across most of the voltage range: the (AA,TT) basepairs lie consistently and clearly below the (GC,AT) basepairs, while the (GG,CC) basepairs are always clearly above. The disparity between the normal and mutation basepairs is especially clear near 0.5V 
Thus, the nanopore device becomes maximally sensitive in distinguishing between normal and mismatched basepairs around this voltage. 

\begin{figure*}[htb]
\centering
\begin{tabular}{cc}
\includegraphics[width=0.42\textwidth]{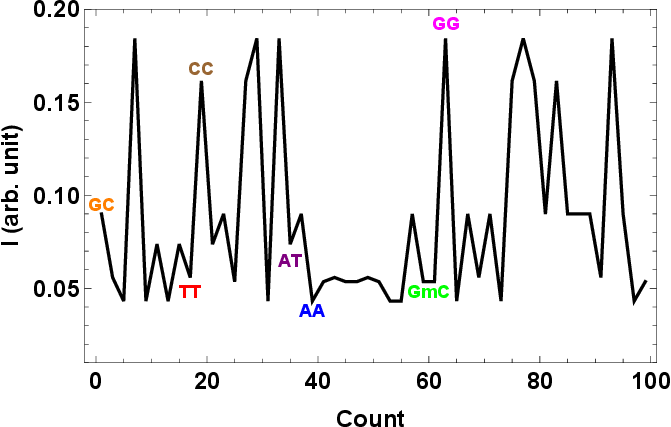}
\includegraphics[width=0.42\textwidth]{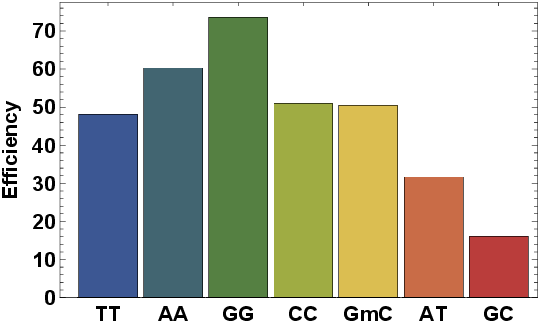}\\
\end{tabular}
\caption{(Color online). Left: Stop-and-Go translocation of a random ATGC ds-DNA chain with mismatches, passed through a strongly coupled nanopore with a fixed bias of 0.5 volts. The characteristic current output for the bases as they pass through the nanopore is recorded, with base-pairs and mismatches indicated using their usual symbols (AT, GC, GG, etc.). Right: Efficiency of the strongly coupled nanopore (see main text for definition), with a maximum of 73.6$\%$ achieved for the GG mismatch.
}
\label{fig7}
\end{figure*}

In Fig.~\ref{fig7}, we illustrate the sequencing application for detecting base pair mismatches along a random ATGC chain in a strongly coupled scenario. A 50-base pair random ATGC chain is translocated through a strongly coupled nanopore following a ``stop and go" strategy in which basepairs are passed through the nanopore cavity one at a time, and the characteristic current signal corresponding to the base pair is recorded. A record of the entire sequence of basepairs, including mismatches, is shown in the left panel of Fig.\ref{fig7}, with a bias of 0.5 Volts maintained during translocation in order to maximize separation between the characteristic currents of different base pairs (as discussed above). The analysis reveals that the transversion type mismatches GG and AA are maximally separated, while AT and TT pairs show minimal separation. The methylated-GC pair (GmC) is distinctly separated from the canonical GC pair, highlighting the device's sensitivity to minute variations in base pairs. Further, to quantify device sensitivity, we introduce an efficiency parameter defined through the ratio $(I_{ch}-I_{0})/I_{0}$, where $I_{ch}$ is the characteristic current for each base pair and $I_0$ is the current output from the bare nanopore. The right panel of Fig.\ref{fig7} shows the values of the efficiency parameter for various basepairs. While the efficiency is higher than 30$\%$ for all basepairs other than GC, the maximum accuracy is achieved for the GG mismatch at 73.6$\%$.

\begin{figure*}[htb]
\centering
\begin{tabular}{cc}
\includegraphics[width=0.42\textwidth]{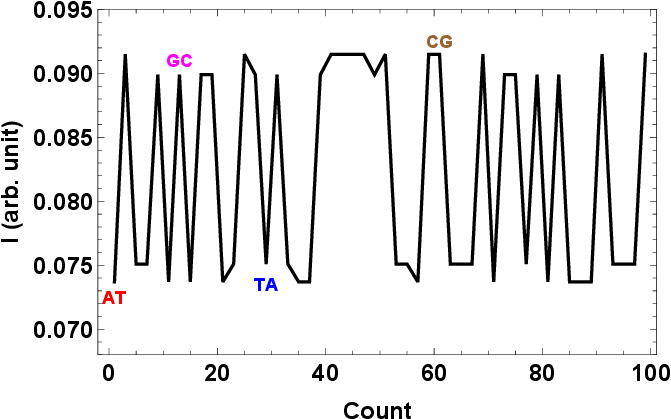}
\includegraphics[width=0.42\textwidth]{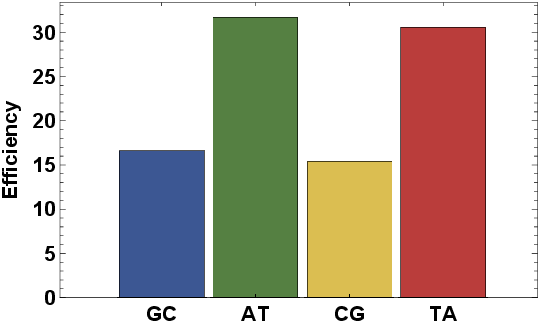}\\
\end{tabular}
\caption{(Color online). Left: Stop-and-Go translocation of a ATGC ds-DNA chain with randomly placed opposite orientations of canonical base pairs, passed through a strongly coupled nanopore with a fixed bias of 0.5 volts. The characteristic current output for the bases as they pass through the nanopore is recorded, with base-pairs and mismatches indicated using their usual symbols (AT, GC, TA and CG). Right: Efficiency of the strongly coupled nanopore (see main text for definition), with a maximum of $\sim 32\%$ achieved for the AT pair.}
\label{fig8}
\end{figure*}

\begin{figure*}[htb]
\centering
\begin{tabular}{cc}
\includegraphics[width=0.42\textwidth]{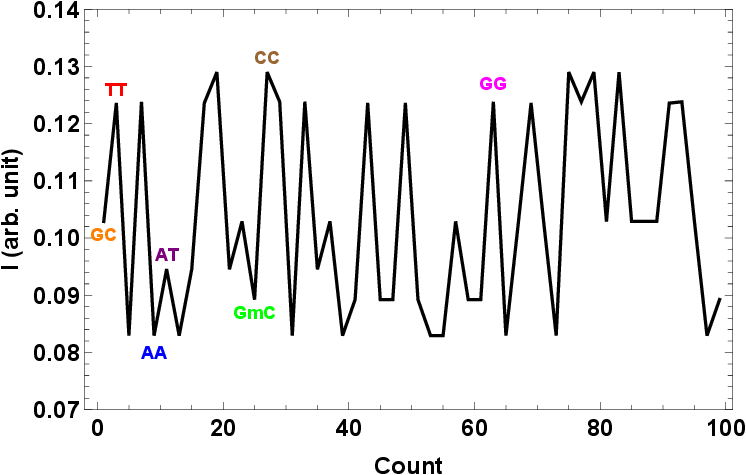}
\includegraphics[width=0.42\textwidth]{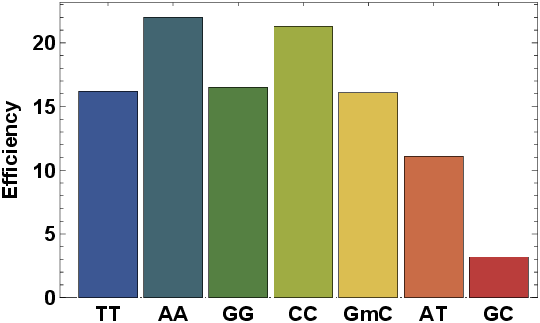}\\
\end{tabular}
\caption{(Color online). Left: Stop-and-Go translocation of a ATGC ds-DNA chain with randomly placed mismatches, passed through a strongly coupled nanopore with a fixed bias of 0.5 volts in the presence of structural fluctuations. The characteristic current output for the bases as they pass through the nanopore is recorded, with base-pairs and mismatches indicated using their usual symbols (AT, GC, GG etc.). Right: Efficiency of the strongly coupled nanopore (see main text for definition), with a maximum of $\sim 22\%$ achieved for the AA mismatched basepair.}
\label{fig9}
\end{figure*}

In order to test further the robustness of such a nanopore device, we conclude by analysing two other cases of interest. First, in Fig.~\ref{fig8}, a different test for the sensitivity of the proposed strongly coupled device is presented. In this case, a random ATGC chain without any mutations is introduced, but with canonical pairs in opposite orientations (e.g., AT and TA, GC and CG). As the relative separation and efficiencies in this case is considerably smaller than that obtained for the case of mutations (Fig.\ref{fig7}), a successful identification of oppositely oriented basepairs from the canonical will need a greater accuracy in measurement of the current through the device. Further, in Fig.~\ref{fig9}, we present the performance of the strongly coupled device in the presence of structural fluctuations during DNA translocation through the nanopore. These fluctuations account for translocation dynamics and base pair orientation relative to the nanopore, including variations at the interface region that can affect electronic couplings. The fluctuations are implemented by randomly varying the coupling ($\tau_2$) within a range 0.1 to 0.5. The effect of these fluctuations on the ability of the device in distinguishing between canonical base pairs and mismatches can be learnt from its efficiency (as defined above). The efficiency values are considerably lowered by translocation-induced fluctuations and changes in orientation: a maximum efficiency of 22$\%$ is obtained for AA type mutations, and a minimum accuracy of 3.2$\%$ for GC pair. This suggests that in order to achieve higher efficiency in such a device, a strategy by which to minimise the role of structural fluctuations will need to be formulated. We leave such investigations for future works.

\section{Concluding Remarks}
The Sanger method~\cite{sanger}, the traditional genome testing technique, remains costly and time-consuming. On the other hand, single-molecule electrical detection techniques are still in development and are currently being tested only on single-stranded DNA. This limits the ability to perform a fast whole genome sequencing for patients with potential genetic disorders. In this study, we tested a graphene nanopore device for sequential determination of double-stranded DNA to detect base-pair mismatches. Specifically, we first studied the characteristic transport properties of two canonical basepairs as well as five different basepair mismatches in a static condition (i.e., by trapping each base pair at the nanopore site). Through current output measurements, we identified the optimal voltage setting for the device to maximize its sensitivity to basepair mismatches. Using this information, we analysed a random double-stranded DNA in order to detect the sequence of basepairs, including any mismatches. The current measurements show that the mismatched basepairs are typically well separated from the canonical ones. An efficiency parameter defined by us quantifies the sensitivity of the device; we find the efficiency for almost all the basepairs to be greater than 30$\%$, and peaks at around 76$\%$. We also tested the proposed device against opposite orientations of canonical pairs (e.g., GC and CG), finding that a higher accuracy in current measurements is needed to distinguish them from their canonical partners. Further, a study of the effects of structural fluctuations revealed a substantially lowered efficiency parameter for the device. This suggests that the proposed device will have to be enhanced with a strategy that minimises such fluctuations. In summary, while our analysis demonstrates the validity of the working principle for such a graphene nanopore sequencing device for the detection of basepair mismatches in a DNA double strand, we have also observed a sensitivity towards operation in near-realistic scenarios that necessitates further improvement of the device. Additionally, the LDOS profiles obtained by us enable detection of nitrogen bases via AFM or STM studies. Thus, our results offer an important first step taken towards the investigation of graphene nanopores that can effectively detect base-pair mismatches, aid in early detection of genetic disorders and improving patient care, and advance the potential for real-time electrical DNA sequencing.

We conclude by discussing a few important points related to our study. First, the current values presented in the figures above are in arbitrary units, because we report model Hamiltonian-based calculations. When the numerical values of constants such as $h$, $e$, and $\hbar$ are included, the currents are on the order of 10~$\mu$A. This is higher than previous reports on ss-DNA sequencing and well above the noise levels of similar devices, which are typically in the nA range~\cite{saha}. Noise studies by Krems \emph{et al.}~\cite{krems} suggest that environmental effects have minimal impact on current distributions and device performance. In addition, currents are measured with nA resolution routinely in experimental bioelectronics nowadays ~\cite{zhuravel2020}; thus, the current ranges reported here are well above the experimental resolution and noise level. However, noise from water and counterions may affect device operation under practical conditions. Functionalizing the nanopore edge atoms can enhance the nucleobase-pore interaction, reducing structural noise and improving the graphene-nucleobase electronic coupling. Functional groups such as hydroxyl~\cite{jeong}, amine, or nitrogen~\cite{saha} can be used for customized noise reduction in electrical DNA sequencing. Engineering the device dimensions can also control the current flow through the nanoribbon edges, addressing sensitivity issues in such devices.

\section*{Acknowledgments}

SK thanks the CSIR, Govt. of India for funding through a research associate fellowship (09/0921(17952)/2024-EMR-I). S.L. thanks the SERB, Govt. of India for funding through MATRICS grant MTR/2021/000141 and Core Research Grant CRG/2021/000852. S.L. also thanks the Anusandhan National Research Foundation, Govt. of India for funding through Advanced Research Grant ANRF/ARG/2025/004414/PS.





\end{document}